# TOWARD LLM-SUPPORTED AUTOMATED ASSESSMENT OF CRITICAL THINKING SUBSKILLS

**Marisa C. Peczuh***
University of Minnesota
peczu001@umn.edu

**Nischal Ashok Kumar***
University of Massachusetts Amherst
nashokkumar@umass.edu

**Ryan Baker**
Adelaide University
ryanshaunbaker@gmail.com

**Blair Lehman**
Brighter Research
blehman@brighter-research.com

**Danielle Eisenberg**
Educational Testing Service (ETS)
deisenberg@ets.org

**Caitlin Mills**
University of Minnesota
cmills@umn.edu

**Payu Wittawatolarn**
University of Massachusetts Amherst
pwittawatola@umass.edu

**Kushaan Naskar**
University of Massachusetts Amherst
knaskar@umass.edu

**Keerthi Chebrolu**
University of Massachusetts Amherst
kchebrolu@umass.edu

**Sudhip Nashi**
University of Massachusetts Amherst
snashi@umass.edu

**Cadence Young**
University of Massachusetts Amherst
cadenceyoung@umass.edu

**Brayden Liu**
University of Massachusetts Amherst
zliu@nutrition.umass.edu

**Sherry Lachman**
Advanced Education Research and Development Fund (AERDF)
slachman@aerdf.org

**Andrew Lan**
University of Massachusetts Amherst
andrewlan@cs.umass.edu

**Joint first authors with equal contribution.*

# ABSTRACT
As the world becomes increasingly saturated with AI-generated content, disinformation, and algorithmic persuasion, critical thinking – the capacity to evaluate evidence, detect unreliable claims, and exercise independent judgment – is becoming a defining human skill. Developing critical thinking skills through timely assessment and feedback is crucial; however, there has not been extensive work in educational data mining on defining, measuring, and supporting critical thinking. In this paper, we investigate the feasibility of measuring "subskills" that underlie critical thinking. We ground our work in an authentic task where students operationalize critical thinking by writing argumentative essays. We developed a coding rubric based on an established skills progression and completed human coding for a corpus of student essays. We then evaluated three distinct approaches to automated scoring: zero-shot prompting, few-shot prompting, and supervised fine-tuning, implemented across three large language models (GPT-5, Llama 3.1 8B, and ModernBERT). Fine-tuning Llama 3.1 8B achieved the best results and demonstrated particular strength on subskills with highly separable proficiency levels with balanced labels across levels, while lower performance was observed for subskills that required detection of subtle distinctions between proficiency levels or imbalanced labels. Our exploratory work represents an initial step toward scalable assessment of critical thinking skills across authentic educational contexts. Future research should continue to combine automated critical thinking assessment with human validation to more accurately detect and measure dynamic, higher-order thinking skills.

# 1. INTRODUCTION
While the ability to think critically has always been an important skill for students to learn, it is increasingly important in today's ever-changing world [14]. Despite widespread recognition of its significance, the development of students' critical thinking has not been a major focus in the educational data mining (EDM) community to date. For example, a search of yearly EDM Conference Proceedings since 2020 revealed only one title and zero abstracts with the term 'critical thinking.'

There are likely several reasons why critical thinking has not been a central focus in educational data mining. Typically, in the EDM community, we must be able to create reasonably accurate ways to assess a skill or construct in order to either proactively or reactively respond to support the learner in a personalized way. Before a construct can be assessed, it must be clearly defined and operationalized, which has been a challenge for critical thinking skills to date. For example, a number of competing definitions and frameworks have articulated distinct visions of what critical thinking is, and how it manifests across contexts [33, 46, 60]. A limitation of many of these prior definitions and frameworks is that they have been vague and difficult to operationalize [4, 5]; others have been criticized for being narrow and over-simplified [57].

A recent definition of critical thinking that is both specific and feasible to operationalize comes from Pasquinelli and colleagues: critical thinking is "the capacity of assessing the epistemic quality of available information and—as a consequence of this assessment—of calibrating one's confidence in order to act upon such information" [46]. Within such a framework, we can think of critical thinking in terms of skills that can be demonstrated, processes needed to effectively use these skills in specific situations, and dispositions that motivate the use of these skills when needed [24]. Critical thinking is, of course, complex and multi-dimensional; successful critical thinking in context calls on skills such as understanding and analyzing information, evaluating evidence, making inferences, articulating one's own arguments, and monitoring one's own thinking process [17, 19, 24, 33, 47]. It's also worth noting that an open question is whether critical thinking skills are domain-general or domain-specific [15, 33, 41, 46]. For this particular submission, we operationalize a set of core "subskills" that are primarily involved in domains that support evaluating and synthesizing information to form new opinions as a foundational example for how to support such modeling.

A key step towards being able to support critical thinking more effectively is being able to assess it rapidly and in context to inform teachers and students [16] as well as adaptive learning platforms. Critical thinking assessments have been developed in multiple formats (e.g., multiple choice, open-ended questions), but these measurements have faced challenges with reliability and validity [33, 37]. Another important limitation to address is the scarcity of assessments that are easily implemented and actionable within educational settings [37]. Developing tools that can detect indicators of critical thinking skills in the moment and automatically, particularly in the context of authentic learning activities, could increase the feasibility for use in adaptive learning and contribute to formative assessment for teachers.

In the current paper, we provide insights into this problem by investigating the feasibility of measuring a set of critical thinking subskills that are core to critical thinking processes (e.g., evaluating evidence, drawing conclusions), rapidly and solely from an artifact produced by an authentic student learning activity, an argumentative essay. We developed a rubric for coding critical thinking subskills based on an existing skills progression, which human raters used to score a set of argumentative essays. We then explored the use of large language models (LLMs) in assessing critical thinking subskills from student-written argumentative essays. We investigated a range of methods, from prompting to fine-tuning, across three different LLMs, leveraging the same rubrics we provided the human coders.

Ultimately, we aim to develop methods that can be used to assess critical thinking within authentic learning activities, across educational activities and contexts. For this paper, we began with a single context (essays) and multiple subskills, providing a proof-of-concept methodology for EDM. We attempted to embrace the inherent complexity of critical thinking, in that it has multiple subskills that exist long a progression rather than being easily operationalized and modeled in a binary manner. Our goal was therefore to test whether we could develop effective models for the present dataset, and which subskills (if any) were amenable to such detection.

# 2. RELATED WORK
## 2.1 Critical Thinking Definitions and Frameworks
As mentioned above, there have been a range of definitions and frameworks for critical thinking [1, 46, 52, 60], but across definitions and frameworks, critical thinking generally involves having the appropriate background knowledge or acquiring information as needed [33], being motivated to engage in critical thinking processes [17, 24], possessing essential critical thinking competencies [19, 57], and being able to use the right competencies for a specific situation in the right fashion [7, 57]. While many frameworks for critical thinking skills exist (see reviews in [1, 33]), and there are many important differences between these frameworks, there are some subskills which the field has come to a consensus about. First, critical thinking involves evaluating the quality and credibility of information (e.g., source, evidence, argument) [17, 19, 30, 33, 47], analyzing the meaning of this information [17, 19, 24, 33, 47], and making inferences, such as how the information might apply in new scenarios [17, 19, 33]. Critical thinking also involves formulating one's own arguments [31], explaining one's own reasoning [17, 19, 33], and forming conclusions [47]. Finally, critical thinkers necessarily self-regulate their processes to continuously monitor and improve [17, 19, 24, 30, 33, 47]. Here we focus on these subskills rather than the overall construct of "critical thinking" in order to target the process involved in a more concrete and definable way.

## 2.2 Measuring Critical Thinking

Regardless of the ongoing debates on definitions and frameworks, there have been several attempts to measure critical thinking skills over the last three decades [8, 16, 33, 37]. The simplest approach to assessing critical thinking has been to give scenarios to students and then ask multiple choice questions [10, 62]. While these assessments can be time efficient, and can have good psychometric properties, there have been concerns about how much information they provide about critical thinking for real-world challenges and authentic activities [37], particularly in a way that can be used to support formative feedback.

Some assessments, such as the Halpern Critical Thinking Assessment [23], have attempted to use open response items and are more relevant to real world situations, where students need to respond authentically in the moment [37]. Still other instruments and rubrics have been used to assess critical thinking in artifacts [16, 20, 37]. For example, the Ennis Weir Critical Thinking Essay Test [18] was designed to assess critical thinking in a more realistic way by asking participants to respond to a fictional argumentative letter. Another example, the Holistic Critical Thinking Scoring Rubric [20], can be used to assess critical thinking as demonstrated in an essay or presentation, whether authentic or designed for the purpose of assessment. However, instruments would ideally give students opportunities to develop skills in authentic learning contexts with quick and accurate AI-supported feedback.

## 2.3 Automated Essay Scoring and Qualitative Coding Approaches

One possible paradigm for how to address this may come from recent work in two areas: automated essay scoring (AES) and automated qualitative coding. AES has been studied for decades, with early approaches relying on surface features such as word counts, lexical diversity, and syntactic patterns to approximate human scoring [2, 48]. Then, neural and transformer-based models shifted the field toward more robust representations of text, enabling AES systems to capture deeper discourse and linguistic patterns [27, 51]. Most recently, large language models (LLMs) have introduced a new approach that does not require training: one can prompt LLMs with hand-crafted scoring rubrics and ask them to estimate the score of an essay (see review in [35]). Similar approaches using prompting of LLMs have been used to assess complex reasoning in shorter text as well [38, 58, 65], building on past work that used previous-generation natural language processing (NLP) techniques [28]. This progression has broadened the scope of automated assessment of text from holistic proficiency measures toward more fine-grained, rubric-aligned scoring tasks.

Recent studies have leveraged AES and automated qualitative coding methods for a number of purposes, and have investigated both their potential and limitations. For example, Xiao et al. [65] introduced a dual-process framework for human-AI collaborative essay scoring, leveraging LLMs to provide both initial predictions and justifications that can support teacher decision-making. Stahl et al. [56] demonstrated that integrating argument mining outputs into transformer-based models can enhance AES and provide a stronger basis for modeling reasoning skills. At the same time, Seßler et al. [53] compared LLM performance with teacher ratings in multidimensional essay scoring, showing that while models achieve promising agreement with human judgments, important gaps remain in transparency and trait sensitivity. Similarly, Liu and colleagues [38] investigated the strengths and weaknesses of different prompting techniques for three different textual data sets, and Simon and colleagues [54] investigated how multi-agent LLM systems could enhance the performance of automated coding for thematic analysis.

Early steps have already been taken to utilize this type of method for aspects of critical thinking. For example, early systems applied previous-generation NLP methods such as Latent Semantic Analysis (LSA) to open-ended responses to capture aspects of argumentation such as the relevance of an argument and the effective use of knowledge [49, 50], demonstrating potential but limited alignment with human judgments. More recently Yeginbergen et al. [66] introduced a framework for counter-argument generation that combines retrieved external evidence with LLM prompting and evaluates outputs using an LLM-as-a-Judge metric, advancing critical thinking assessment toward more evidence-grounded and generative application. These approaches show the possibility of using natural language processing to study and assess critical thinking in authentic or similar artifacts, but have not yet been successful at assessing critical thinking multi-dimensionally, in a fine-grained fashion, and across educational activities and contexts. Thus, while our task is not AES in the conventional sense of scoring overall writing quality, it is closely related methodologically: we similarly evaluate whether automated systems can assign human-aligned rubric scores to student essays, but we shift the scoring target from writing quality to critical thinking subskills.

# 3. METHOD

## 3.1 Dataset

We used a portion of the PERSUADE 2.0 corpus [11] that was made publicly available as part of the Feedback Prize - Predicting Effective Arguments competition [21] in our work. The PERSUADE 2.0 corpus includes more than 25,000 argumentative essays that were written by 6th to 12th grade students in the United States, for 15 prompts on two writing tasks: independent and source-based writing. The training data that was released as part of the competition included 4,115 essays from the original corpus. We did not have access to the essay prompt, writing task, or source materials (in the case of source-based writing) in the publicly available dataset, and therefore did not use this information in our human or LLM coding. We selected this dataset as an initial use case for the development of a critical thinking model since argumentative essays (and argumentation more generally) often include evidence of critical thinking processes (e.g., providing evidence for a claim, evaluating alternative viewpoints through counterarguments).

## 3.2 Critical Thinking Rubric

We developed a rubric for coding critical thinking based on an early version of the Critical Thinking Progression for High School Students developed by the Skills for the Future (SFF) initiative, a partnership between ETS and The Carnegie Foundation for the Advancement of Teaching [55].[1] The progression defined critical thinking as "the skill and disposition to actively seek and evaluate information and construct evidence-based arguments to reach well-founded conclusions or informed decisions, including recognizing and applying sound logical reasoning. It plays an important role in learning activities, such as evaluating multiple sources, identifying assumptions and reasoning flaws, and making sound arguments. Developing strong critical thinking skills enables students to tackle complex challenges more effectively and achieve greater learning outcomes" [55].

The skills progression included four skills that were further specified into subskills. The first skill, *Information Seeking*, was not included, as the focal dataset for the present work was completed essays. The remaining three skills were retained for the present work and are shown in Table 1. While these three skills were identified from the SFF initiative, we grounded the subskills in our literature review. For example, many definitions of critical thinking include the ability to determine the trustworthiness of information [17, 19, 30, 33, 47], otherwise known as *Evaluating Evidence Strength* in SFF.

Three proficiency levels were included in the original progression: 'Emerging' (representing lowest level of proficiency, where students are "beginning to identify and understand key aspects of critical thinking"), 'Expanding' ("developing [critical thinking] skills"), and 'Exemplifying' ("demonstrating proficiency" and mastery of critical thinking skills). After initially exploring the feasibility of using the progression for our desired purposes quantitatively and qualitatively, we decided to add two proficiency levels to the rubric that demonstrated lower proficiencies than the 'Emerging' level: 'Below Emerging' (where essays did not yet meet the threshold for 'Emerging' criteria) and 'Not Applicable' (the lowest proficiency level, where "certain critical thinking elements are absent from student contribution," which inhibits a proficiency rating").

We also adjusted the criteria across subskills for achieving various proficiency levels. First, we adjusted the progression to increase clarity in coding and reduce cognitive load of coders, such as collapsing across two sections for each subskill, using parallel wording across proficiency levels, and making relative rather than numerical statements (e.g., "Uses more opinions than facts," "Uses more logical fallacies than evidence"). Second, rather than modifying the progression directly, we provided additional context, such as by broadening the criteria for citing sources in *Synthesizing Multiple Sources* (e.g., direct quotes, reference to author, title of a text) and re-defining evidence in *Evaluating Evidence Strength* (e.g., anything used to support a claim, which can include both facts and opinions/anecdotes).

Third, we made distinctions between certain subskills (i.e., to avoid essays being scored at lower proficiency levels in multiple subskills for the same mistake) while also connecting subskills to align greater critical thinking with better quality arguments. For example, it was necessary for writers to have a clear conclusion (*Drawing Conclusions*) to connect evidence to the conclusion (*Evaluating Evidence Strength*) and evaluate the presence and quality of counterarguments (*Using Counterarguments*). Ultimately, the final rubric included five proficiency levels (from 'Not Applicable' to 'Exemplifying'), three skills and six subskills, and various definitions and resources to support the coding process.

## 3.3 Human Coding

### *3.3.1 Coders and Codebook*

The original coders were four graduate students at a large university in the U.S. (Coder Group 1). All coders had Master's degrees and were advanced students (fourth through sixth years) in an educational psychology PhD program. Their engagement included three meetings to achieve sufficient reliability and independent coding of 50 essays (200 essays total). The codebook development primarily focused on adapting the original Critical Thinking Progression into a rubric that could be applied to coding a specific student-generated artifact (e.g., essay), which is detailed previously in Section 3.2. However, the need for supplementary information was identified during the codebook development and initial training meetings with coders. For example, clarification was needed for *Synthesizing Multiple Sources* as to what qualified as a citation since the essays were written by grade 6–12 students and how to discriminate between summarizing and integrating multiple sources.

A second group of coders (Coder Group 2) was recruited to expand our sample of human-coded essays. Coder Group 2 included three graduate students, two of which were from another large university in the U.S. and one from a large university in Australia. All coders had Master's degrees, and one was in the final year of another Master's program. All coders had extensive prior experience with qualitative coding. Their engagement included three meetings to achieve sufficient reliability and independent coding of 100 essays (300 essays total). The final codebook that was used for Coder Group 1 was used for Coder Group 2 with no additional modifications.

### *3.3.2 Reliability*

Given the complexity of critical thinking as a construct and our overarching goal for this work as usable in practical (i.e., build a critical thinking model that can be used for similar purposes to iteratively evaluate AI tools), we decided on a level of reliability consistent with our preliminary work. Therefore, we did not use the typical thresholds for evaluation criteria for achieving interrater reliability for Coder Group 1 (e.g., Krippendorff's $\alpha \geq .8$) [29, 40, 63]. Instead, we set our threshold to a Krippendorff's Alpha greater than or equal to .6 as our criteria for sufficient interrater reliability. We adopted Krippendorff's Alpha as we had four coders who during the process of achieving reliability provided ordinal codes for all subskills in all essays [25]. The coders scored each subskill holistically by providing a single rating for each subskill for each essay. We used the K-Alpha Calculator [40] to compute Krippendorff's Alpha for each subskill in each attempt to achieve sufficient interrater reliability, described next.

[1] The current version of the Skills for the Future Progressions were made public in January 2026 (https://www.ets.org/newsroom/ets-carnegie-release-skills-progressions.html). They have four levels intended for understanding skill acquisition during secondary school. The project continues to iterate on ways to portray the ways in which skill development is not linear and is currently working to validate these progressions with educators and students.

**Table 1. Critical thinking subskills**

| Skill | Subskill | Definition | Example |
|---|---|---|---|
| 2: Information Analysis | 2.1 Synthesizing Multiple Sources | Effectively synthesizes multiple pieces of information | Summarizes information from multiple sources (with citations) but does not integrate information |
| | 2.2 Evaluating Evidence Strength | Evaluates the strength and relevance of evidence used to form a conclusion | Presents evidence and links evidence to a specific conclusion, but does not evaluate the relevance or strength of the evidence for the argument generated |
| 3: Argument Generation | 3.1 Using Counterarguments | Effectively addresses counterarguments | Acknowledges specific opposing viewpoint(s), counterargument(s), or qualifier(s) |
| | 3.2 Using Facts and Opinions | Relies on data and/or facts over opinions | Uses facts and opinions about equally to support claim(s) and/or arguments |
| 4: Logical Reasoning | 4.1 Drawing Conclusions | Draws specific conclusions | Draws a specific conclusion to analyze simple and straight-forward relationships/argumentations |
| | 4.2 Using Logical Fallacies | Recognizes and avoids logical fallacies | Uses logical fallacies and evidence about equally when generating arguments |

Two rounds of coding were required to achieve sufficient interrater reliability for Coder Group 1, and each round adopted the same approach. Coders were asked to code 10 essays for all six subskills (see Table 1). Prior to the first round of coding, coders engaged in a 90-minute training session. During this session, they also coded an essay collaboratively as a group. The coders engaged in a 60-minute disagreement resolution session that reviewed disagreements in round one codes as well as identified instances in which rubric revisions were necessary (see Section 3.2). The coders then completed a second round of coding with new essays that were coded for all six subskills. The second round of coding resulted in sufficient reliability for all subskills (Table 2), which exceeded our pre-defined threshold for sufficient reliability ($\alpha \geq .60$) and generally achieved a satisfactory level of agreement [40]. The one exception was *Using Facts and Opinions*, which did not reach our pre-defined threshold for sufficient reliability. We chose to proceed with coding of *Using Facts and Opinions* as the discussion of coder disagreements revealed that the disagreements generally stemmed from the coders not having access to the source material provided to students when writing the essays to determine whether statements were fact or opinion. Since our goal was to explore whether it was possible to engage in the process of defining, coding, and predicting critical thinking indicators (rather than develop the most "effective" model), we concluded the interrater reliability had reached a sufficient level to proceed.

**Table 2. Reliability for human coder Group 1 and Group 2**

| Skill | Coder Group 1 Krippendorf's Alpha | Coder Group 2 Weighted Kappa *M (Min-Max)* |
|---|---|---|
| Synthesizing Multiple Sources | 1.00 | 0.87 (0.79-0.91) |
| Evaluating Evidence Strength | 0.68 | 0.74 (0.66-0.78) |
| Using Counterarguments | 0.70 | 0.72 (0.65-0.78) |
| Using Facts and Opinions | 0.57 | 0.82 (0.80-0.83) |
| Drawing Conclusions | 0.75 | 0.69 (0.55-0.80) |
| Using Logical Fallacies | 0.79 | 0.79 (0.69-0.84) |

Although reliability was achieved after the second round, coders met for another 60-minute session, where additional revisions to the rubric were made based on coder disagreement and feedback (see Section 3.2). Then, each coder was assigned 50 essays to independently code for each of the six subskills, which resulted in 200 human-coded essays (5% of total essay sample).

Based on initial model performance, we determined that additional human-coded essays were necessary. Coder Group 2 engaged in the same 90-minute training session as Coder Group 1 and then proceeded to achieve reliability. Given that the goal was to increase the sample of human-coded essays, Coder Group 2 did not seek to achieve sufficient reliability within their group but rather to achieve sufficient reliability with Coder Group 1. This meant that essays coded by Coder Group 1 were leveraged for Coder Group 2 reliability along with the final version of the rubric and that reliability was computed between each coder from Coder Group 2 with Coder Group 1 codes. Weighted Kappas (Quadratic) were used for evaluating interrater reliability with the same threshold used for Coder Group 1 ($\alpha \geq 0.6$). Two rounds of coding (with a 60-minute meeting after each round) were required to achieve sufficient interrater reliability for each Coder Group 2 coder for each subskill (Table 2), except for one coder slightly below our threshold for *Drawing Conclusions* (*Kappa* = .55). Coders were then each assigned 100 essays to independently code for each subskill, which resulted in an additional 300 essays (7% of total essay sample).

The total human-coded essay sample included 500 essays (12% of total essay sample) that were used to evaluate the performance of the LLM critical thinking model. While we recognize this is a modest number of human-coded essays given the nuanced and multifaceted nature of critical thinking and its operationalization, the number is consistent with the overall exploratory goal of this work. Additionally, the sample size is in the range commonly used in other research involving evaluations of whether LLMs agree with human codes [6, 9, 38, 43, 45].

## 3.4 LLM-based Assessment

We evaluated two complementary approaches for LLM-based critical thinking assessment: (1) prompting proprietary LLMs, specifically OpenAI's GPT-5, and (2) fine-tuning two open-source models for classification purposes, Llama 3.1 8B [22] and ModernBERT [61], on our student essay dataset.

### *3.4.1 Prompting and Fine-tuning*

We prompted GPT-5, a state-of-the-art LLM at the time of this work, with two settings: *zero-shot* [34] and *few-shot* [65]. In both settings, we instructed the LLM to assume the role of a calibrated educational scorer, tasked with classifying essays on fine-grained critical thinking subskills according to our structured rubric. The prompt included: (a) the general definitions of the five proficiency levels (0 = 'Not Applicable' through 4 = 'Exemplifying'), and (b) the subskill definition and subskill-specific rubric descriptors. In the **zero-shot condition,** each essay was paired with a single target subskill. We asked the model to output both a proficiency label (0–4) and a concise justification aligned with the rubric. All 500 essays in the dataset were evaluated under this setting. In the **few-shot condition,** the prompt included the same instructions as above, with in-context examples for each proficiency level for the target subskill. We experimented with one, three, and five randomly selected in-context examples per proficiency level for each target subskill. To avoid data leakage, we excluded essays used as examples from the evaluation and tested the model on the remaining essays. For GPT-5, we used the default *reasoning effort* setting to balance performance and efficiency, and set the maximum output length to 3000 tokens to accommodate the model's reasoning traces and justification.

To complement our prompting experiments, we fine-tuned two open-source LLMs, Llama-3.1 8B Instruct, a decoder-only model with strong performance on several popular natural language processing benchmarks and ModernBERT-base, an encoder-only transformer, chosen for its lightweight design (149M parameters) and strong performance-size tradeoff on classification tasks. ModernBERT improves upon its predecessor BERT [13] with training on much larger data and an 8k-token context window, making it well-suited for efficiently handling student essays.

For fine-tuning Llama, we constructed the input prompt by including the subskill name, definition and subskill-specific rubric descriptors corresponding to the five proficiency levels followed by the essay text to be classified. The output of the model is the proficiency level label text itself ('Not Applicable,' 'Below Emerging,' 'Emerging,' 'Expanding,' 'Exemplifying'). We fine-tuned the model using PEFT with LoRA adapters (rank=32, α=64, dropout=0.05), updating only adapter weights [26]. Training used a chat-style input-output format, optimized with AdamW for 3 epochs (batch size 4, gradient accumulation 4) with a 1,536-token sequence length. We selected the checkpoint with the best validation loss on a held-out validation set for inference.

For fine-tuning ModernBERT, like Llama, we construct the input prompt by including the subskill name, definition and subskill-specific rubric descriptors corresponding to the five proficiency levels. The output of the model is a single proficiency label (0–4). We added a lightweight classification head over the [CLS] representation and trained with AdamW and focal loss [36] to mitigate class imbalance for 6 epochs, using early stopping and a maximum input length of 1,536 tokens.

For Llama fine-tuning, in addition to outputting the label text, we experimented with a variant, where the model first outputs a short justification and then outputs the label text in order to elicit a chain-of-thought behavior [64] in the LLM to help in the reasoning process to generate the label [65]. Since our dataset does not contain gold justifications written by the human coders, we synthetically generate them by prompting OpenAI's GPT-4.1 to generate a sentence justifying the human label for a particular subskill for a given essay. We call this fine-tuning method "Llama (with justification)". In contrast, the default variant outputs only the proficiency label text (without generating a justification), which we refer to as "Llama (without justification)".

Additionally, to study the role of the subskill-specific rubric descriptors corresponding to the five proficiency levels in the input prompt, we consider two input prompt variants for both Llama and ModernBERT. In the with-rubric variant, the input prompt includes the subskill name, definition, and all rubric descriptors, along with the essay text. In the without-rubric variant, we remove the rubric descriptors and include only the subskill name and its definition, and the essay text. We refer to these variants as "Llama (with rubric)/ ModernBERT (with rubric)" and "Llama (without rubric)/ ModernBERT (without rubric)", respectively.

We evaluated two data-splitting regimes: 1) **Essay-based split**: essays divided into 70/10/20 percentages for the train/validation/test sets, respectively, with subskills constant, and results averaged across five random seeds. 2) **Subskill-based split**: one subskill held out entirely for testing; the other two used for training and validation with a 90/10 essay split. The first splitting setup is standard and evaluates LLMs' scoring ability when given on-task training data, i.e., human scored essays on each subskill. The second splitting setup tests generalization to new subskills, without using any human scores for previously unseen subskills.

### *3.4.2 Evaluation Metrics*

To compare the performance of LLMs against human-annotated data, we evaluated performance across several metrics, each chosen to capture a different perspective on model behavior. **Accuracy** measures the proportion of exact matches between predicted and true labels. It offers an intuitive sense of overall correctness, but does not capture the severity of errors when predictions fall close to, but not exactly on, the target proficiency level. It also does not take class imbalance into consideration. Root Mean Squared Error (**RMSE**) treats rubric levels as points on an ordinal scale and penalizes deviations according to their distance. By squaring errors before averaging, larger deviations between the predicted score and the actual score are weighted more heavily, making RMSE suitable for ordinal classification tasks. **Macro F1** calculates the F1 (harmonic mean of precision and recall) score independently for each class and then averages across classes, giving equal importance to rare and frequent proficiency levels. This setup highlights whether a model can handle minority categories, such as 'Exemplifying,' as effectively as majority ones. **Weighted F1** also averages per-class F1 scores but weights them by class frequency, producing a summary that reflects the overall distribution of labels. This setup makes it more representative of performance in imbalanced datasets where some categories dominate. **Krippendorff's α** treats the model as a second annotator and quantifies agreement with human labels beyond chance. Since it is defined for ordinal scales, α accounts for the graded nature of rubric levels, providing a more nuanced measure of reliability than accuracy alone.

It is also important to note that there is not an expectation of near perfect performance in this case. There are no "correct" or "incorrect" ways to critically think, which is the case even for human coders. Therefore, it is not expected that an LLM would detect critical thinking subskills even better than humans. Rather, it is more likely that the LLM's would perform similarly to the detection of other complex cognitive constructs in the EDM field, such as frustration, engagement, and mind wandering [32, 42, 44]. To strengthen the trustworthiness of our results, we therefore evaluate our approach using multiple complementary metrics and, for the essay-based split, report averages across five random train/validation/test splits. This setup reduces reliance on any single metric or data split and supports interpreting the results as stable trends within this exploratory dataset.

**Table 3. Distribution of critical thinking proficiency levels by subskill**

| Skill | Not Applicable | Below Emerging | Emerging | Expanding | Exemplifying |
|---|---|---|---|---|---|
| Synthesizing Multiple Sources | 287 | 168 | 19 | 18 | 8 |
| Evaluating Evidence Strength | 48 | 73 | 363 | 14 | 2 |
| Using Counterarguments | 231 | 113 | 62 | 77 | 17 |
| Using Facts and Opinions | 58 | 140 | 122 | 172 | 8 |
| Drawing Conclusions | – | 66 | 299 | 113 | 22 |
| Using Logical Fallacies | 15 | 121 | 132 | 230 | 2 |

# 4. RESULTS

## 4.1 Human Coding

The independent human coding resulted in 500 essays with proficiency level codes for each subskill. Table 3 shows the distribution of essays coded for each proficiency level across subskills.

For *Synthesizing Multiple Sources*, the most prominent proficiency level was 'Not Applicable,' which represented that the essays did not cite any sources, followed by 'Below Emerging,' which signifies that the essays only cited one source. Thus, for the majority of essays (91%), there was no opportunity to evaluate the degree to which students synthesized multiple sources, since multiple sources were not cited. For *Evaluating Evidence Strength*, 'Emerging' was the most prominent proficiency level (73%), which signifies that the essays presented evidence that was linked to specific conclusions, but did not evaluate the strength or relevance of the evidence.

For *Using Counterarguments* 'Not Applicable' was the most prominent proficiency level (46%) and suggests that the essays did not include any counterarguments. The next most prominent proficiency level was 'Below Emerging (23%) which included essays that only acknowledged vague counterarguments (e.g., others may disagree; 20%). 'Expanding' was the most prominent proficiency level for *Using Facts and Opinions* (34%), which applies to essays that used more facts than opinions to support the stated claims or arguments, followed by 'Below Emerging' (28%), which applies to essays that used more opinions than facts to support the stated claims or arguments. This finding may be explained by the variation in essay prompts and writing tasks as a review of the essays revealed that students were likely prompted in some to give their opinion, whereas in others they were prompted to leverage facts in provided source materials.

'Emerging' was the most prominent proficiency level for *Drawing Conclusions* (60%), followed by 'Expanding' (23%). This suggests that essays were typically able to draw a specific conclusion but were more likely to analyze simple, rather than complex, relationships. For *Using Logical Fallacies*, the most prominent proficiency level was 'Expanding' (46%) with 'Emerging' as the next most prominent (26%). Essays in the present sample tended to include evidence at an equal or greater amount than logical fallacies when generating arguments.

It is also noteworthy that across subskills the 'Exemplifying' proficiency level was extremely rare (<1% – 4%). This may be due to the essays being authored by students between grades 6 and 12 while the learning progression was developed specifically for high school students. The uneven distribution of proficiency levels, especially the rarity of 'Exemplifying' responses, also highlights a broader challenge for automated critical thinking assessment: models must be validated on additional datasets with different student populations, tasks, and proficiency distributions before claims can be generalized beyond the present setting.

## 4.2 LLM Scoring Results

### *4.2.1 Subskill Scoring*

In the first experiment, we compared the performance of all three methods (zero-shot prompting, few-shot prompting, fine-tuning) and all three base LLMs (GPT-5, Llama, ModernBERT) on each subskill. We show the overall results in Table 4, averaged across test sets in five random seed splits of the dataset. Overall, fine-tuning Llama yields the strongest performance. The best variant is Llama (with rubric, without justification), which conditions on the subskill-specific rubric descriptors and predicts only the label. Importantly, Llama is the only model that crosses the $\alpha \geq .6$ threshold we originally adopted as sufficient reliability for human coders. This result indicates that fine-tuning a capable base LLM on our dataset is necessary to reach satisfactory performance, considering what is deemed acceptable for human interrater agreement.

Fine-tuning ModernBERT (with rubric) is not as effective as fine-tuning Llama: it underperforms Llama across all metrics with a notable gap of 4% and 6.5% in accuracy and Krippendorff's α respectively. For ModernBERT, rubric descriptors have limited impact: ModernBERT (without rubric) is only .8% lower in accuracy, and ModernBERT (with rubric) is only 1% lower in Krippendorff's α, suggesting the model relies primarily on the essay text during fine-tuning. We also found that ModernBERT is effective at capturing majority labels but struggles with less-frequent, minority labels, reflecting the challenges of fine-tuning on an imbalanced dataset even with the focal loss. These results highlight ModernBERT's limited model size (149M parameters compared to 8B parameters), making it less capable than Llama. Also, the decoder-only nature of Llama means that it generates the label text directly, instead of using a separate classification head for each subskill for ModernBERT. Therefore, it is easier for Llama to make the most use of textual rubric definitions for subskills.

**Table 4. Scoring results on the essay-based split, averaged across test sets from five random seeds**

| Model | Method | Accuracy | RMSE | F1 (Macro) | F1 (Weighted) | Krippendorff's α |
|---|---|---|---|---|---|---|
| Llama 8B | Fine-Tune | .604 ± .012 | .933 ± .043 | .489 ± .022 | .599 ± .016 | .606 ± .032 |
| ModernBERT | Fine-Tune | .563 ± .021 | 1.072 ± .066 | .448 ± .019 | .552 ± .025 | .541 ± .044 |
| GPT-5 | Zero-Shot | .543 ± .013 | 1.094 ± .045 | .443 ± .011 | .541 ± .013 | .487 ± .031 |
| | Few-Shot | .532 ± .013 | 1.079 ± .031 | .457 ± .023 | .534 ± .015 | .542 ± .016 |

**Table 5. Results showing Llama fine-tuning variants using rubrics and justification**

| Rubric | Justification | Accuracy | RMSE | F1 (Macro) | F1 (Weighted) | Krippendorff's α |
|---|---|---|---|---|---|---|
| With Rubric | Without Justification | .604 ± .012 | .933 ± .043 | .489 ± .022 | .599 ± .016 | .606 ± .032 |
| | With Justification | .573 ± .014 | .974 ± .051 | .455 ± .011 | .562 ± .018 | .588 ± .022 |
| Without Rubric | Without Justification | .591 ± .015 | .942 ± .036 | .481 ± .025 | .586 ± .019 | .599 ± .034 |
| | With Justification | .576 ± .004 | .983 ± .035 | .462 ± .011 | .563 ± .006 | .584 ± .011 |

Prompting GPT-5 with few-shot (with three in-context examples) yields better performance in comparison with zero-shot (~5.5%) in terms of Krippendorff's α with minimal differences across other metrics. This result shows that the in-context examples help in classifying the minority labels, improving the overall agreement with humans. Few-shot prompting with different numbers of in-context examples (one, three, and five) yields very similar results, with three in-context examples being slightly better (by 1% on accuracy and α) showing that increasing the number of in-context examples does not yield significant gains. The modest gains from few-shot prompting for GPT-5 also highlight its strong zero-shot ability: the model can adapt to the rubric with little additional context.

Comparing fine-tuning to prompting, we see that ModernBERT, despite being a smaller model, matches GPT-5 few-shot in terms of Krippendorff's α, while Llama fine-tuning outperforms it by 6%. Overall, fine-tuning smaller, open-source models outperforms prompting larger, proprietary start-of-the-art LLMs. These results highlight the importance of training with on-task data, since LLMs are not pre-trained on our specific task of critical thinking assessment. There are also additional benefits in cost and security: few-shot prompting has significantly higher cost (≈ $40 for scoring 500 essays with GPT-5 via the OpenAI API) and raises concerns about student data privacy.

### *4.2.2 Llama 3.1 8B Results*

Table 5 shows results on different variants of Llama fine-tuning using rubrics and justifications. We see that including the subskill-specific rubric descriptors in the input prompt yields only slight improvement than not including it (by 1.3% in accuracy and .7% in Krippendorff's α for the Llama [without justification] variant). This result suggests that the model attends mostly to the essay text and the subskill name when making its classification, while the rubric information does not play a very significant role. This result can perhaps be explained by the data split, where scored examples of the same subskills are already seen during training, resulting in on-task training data. However, such rubrics may be useful when we generalize to unseen subskills, which we study in Section 4.2.3. We also see that not including the justification performs slightly better than including it in the output (by 3.1% in accuracy and 1.8% in Krippendorff's α for the Llama [with rubric] variant). This result suggests that the justifications might be noisy: they are not written by humans, which may add additional distracting information during the model's decision-making process.

Table 6 shows the results for our best-performing variant, Llama (with rubric, without justification) broken down across different subskills. We see that both the subskills in *Subskill 2 - Information Analysis* achieve the highest accuracy (around .78) compared to other subskills. We note that this result is likely due to the label distribution for both subskills being highly skewed, with the model mostly predicting the majority label. However, we do see a significant difference in the Krippendorff's α scores across subskills, with *Synthesizing Multiple Sources* having a Krippendorff's α of .74, showing the fine-tuned LLM as quite consistent with human judgement. On the contrary, *Evaluating Evidence Strength* has a low Krippendorff's α score of only .23. This difference is because the model always predicts the majority 'Emerging' label for *Evaluating Evidence Strength*, ignoring the 'Not Applicable' and 'Below Emerging' labels that are not seen much during training.

We also see that both subskills under *Subskill 3 - Argument Generation* achieve the lowest accuracy among the other subskills. This result can be explained by the balanced label distribution across the proficiency levels, where subtle differences in essays led to differences in labels, making it hard for LLMs to learn during training. Despite the lower accuracy, we see that *Using Facts and Opinions* achieves a Krippendorff's α of .55, which is close to the threshold we set for interrater agreement between human coders.

**Table 6. Subskill-wise results for the best Llama fine-tuning variant**

| Subskill | Accuracy | RMSE | F1 (Macro) | F1 (Weighted) | Krippendorff's α |
|---|---|---|---|---|---|
| Synthesizing Multiple Sources | .780 ± .037 | .638 ± .108 | .444 ± .061 | .761 ± .040 | .740 ± .063 |
| Evaluating Evidence Strength | .768 ± .026 | .707 ± .097 | .303 ± .043 | .711 ± .033 | .234 ± .126 |
| Using Counterarguments | .471 ± .042 | 1.408 ± .170 | .236 ± .050 | .416 ± .065 | .315 ± .148 |
| Using Facts and Opinions | .487 ± .024 | .958 ± .045 | .325 ± .044 | .440 ± .044 | .555 ± .032 |
| Drawing Conclusions | .613 ± .033 | .672 ± .033 | .361 ± .057 | .555 ± .053 | .382 ± .091 |
| Using Logical Fallacies | .503 ± .027 | .967 ± .032 | .348 ± .036 | .450 ± .061 | .411 ± .049 |

Finally, both subskills in *Subskill 4 - Logical Reasoning* have higher accuracy than *Subskill 3 - Argument Generation* but lower than those of *Subskill 2 - Information Analysis*. In particular, *Using Logical Fallacies* has an accuracy of only .50, compared to .61 for *Drawing Conclusions*. This result underscores the challenge in identifying different types of logical fallacies and differentiating them from the evidence, since the fallacies are often subtle and rarely labeled explicitly.

Taken together, these results suggest that fine-tuning Llama performs best in contexts where different proficiency levels are separable and well-represented. Performance significantly drops when different proficiency levels have minor differences or the labeled data is dominated by one level. Therefore, it is important to evaluate models with metrics beyond accuracy before considering potential classroom use.

### *4.2.3 Generalizing to Unseen Subskills*

In this experiment, we evaluate whether a Llama model fine-tuned on labeled data on some subskills can generalize to other subskills that are previously unseen during training. Intuitively, since the definition of subskills and scoring rubrics defining each proficiency level are used as input, a trained model may acquire some knowledge on how to evaluate students' critical thinking as a whole, as well as gain knowledge on the operational definitions of the different proficiency levels. Therefore, the trained model may perform better in the cold-start setting, where human scores are not readily available for previously unseen subskills [64].

Table 7 shows results on the subskill-based data split where we see a clear performance drop compared to the essay-based split. Among the variants, Llama (with rubric, with justification) performs best in this setting, achieving Krippendorff's α around .42. However, the without-rubric variants, either with or without justification, achieve α values near zero. This result suggests that rubric descriptors are essential for cross-subskill generalization: Including the rubric provides the model with the context it needs to map an unseen subskill to proficiency level labels. This observation is in stark contrast to the essay-based split, where rubrics are not essential since the data split does not test generalizability to new subskills. For the rubric-based variants, adding justifications improves generalization: Llama (with rubric, with justification) outperforms Llama (with rubric, without justification) by 10% in accuracy and 6% in Krippendorff's α. This result differs from the essay-based split, where justifications provided little benefit. This observation suggests that generating justifications improves the model's reasoning by encouraging it to explain its label using rubric descriptors and evidence from the essay, which matters more when the target subskill is unseen during training.

Unfortunately, accuracy is low for some subskills (between 0.39 and 0.56), and Krippendorff's α falls below the reliability threshold we set for human annotators. These results suggest that models struggle to generalize when the target subskill is unseen during training. This observation underscores the importance of subskill-specific context for guiding classification. In particular, models rely on the rubric descriptors for each subskill; removing this information during training substantially degrades performance. These results suggest that effective scoring of new subskills will require subskill-specific labeled data.

Despite this difficulty in generalization, Information Analysis achieves the highest accuracy among the unseen subskills (.56). This result likely follows because the subskill has simpler classification boundaries and higher human interrater agreement, which makes the dominant label easier for the model to identify and predict consistently. Argument Generation achieves the highest Krippendorff's α (.41) but lower accuracy and higher RMSE. This result suggests that the model struggles to separate adjacent labels when the label distribution is more balanced, leading to confusion among intermediate levels. Logical Reasoning achieves the lowest Krippendorff's α (.30). This result likely reflects the complexity of the subskill, especially *Using Logical Fallacies*, which requires recognizing specific fallacy types and distinguishing them from evidence.

Taken together, these findings highlight the difficulty of generalizing across subskills without on-task training data. These findings also show some promise in generalization, with accuracy comparable to the essay-based split for a few subskills and fair Krippendorff's α scores across all subskills. Compared to GPT-5 zero-shot, which does not rely on training data for unseen subskills either since it is a prompting method, Llama fine-tuning performs slightly worse, by 8.6% overall in accuracy and 6.5% overall in Krippendorff's α. This drop in performance likely follows from subskill definitions and rubric criteria being sufficiently distinct, making extrapolation to other subskills difficult. Consequently, robust scoring of unseen subskills remains challenging and may require some subskill-specific supervision or LLM fine-tuning techniques that encode subskill-agnostic features.

### *4.2.4 Error Analysis*

We conducted an error analysis of the best-performing method, fine-tuning Llama (with rubric, with justification) on *Using Logical Fallacies* (Subskill 4.2). The most common misclassifications are 'Expanding' classified as 'Emerging' (80), 'Emerging' classified as 'Below Emerging' (53), and 'Emerging' classified as 'Expanding' (46). We qualitatively examined three representative examples, one for each misclassification type.

**Table 7. Scoring results on the subskill-based split for Llama fine-tuning**

| Test Subskill | Accuracy | RMSE | F1 (Macro) | F1 (Weighted) | Krippendorff's α |
|---|---|---|---|---|---|
| Information Analysis | .564 | 1.221 | .358 | .549 | .379 |
| Argument Generation | .389 | 1.357 | .293 | .356 | .418 |
| Logical Reasoning | .418 | 1.066 | .310 | .406 | .302 |
| Overall | .457 | 1.22 | .392 | .448 | .422 |

The first example discusses electoral rules. The essay includes concrete evidence such as “The Electoral College consists of 538 electors… 270 electoral votes” and “Because of the winner-take-all system… the candidates don’t spend much time.” The model instead foregrounds fairness appeals such as “It’s not fair Texas gets 38… New York only gets 3,” and similar overgeneralizations as logical fallacies and minimizes evidence, leading it to predict ‘Emerging’ instead of ‘Expanding’. The second example discusses classroom emotion-detection software. The essay includes technical evidence describing the Facial Action Coding System, such as “orbicularis makes crow’s feet around your eyes” and “the mouth is stretched sideways using the zygomatic major and a different muscle called the risorius.” The model instead emphasizes slippery-slope and speculative claims (e.g., “more and more ads… you might even get a virus,” “watching you all day, everyday”), rather than “concrete evidence,” leading it to predict ‘Below Emerging’ instead of ‘Emerging’. The third example discusses Venus exploration. The essay includes research-like statements such as “far more extreme than anything humans encounter on Earth” and “97 percent carbon dioxide blankets Venus.” The model instead gives disproportionate weight to the volume of citations while not recognizing contradictions in the essay’s conclusion, leading it to predict ‘Expanding’ instead of ‘Emerging’.

Taken together, these results indicate that most misclassifications occur between adjacent proficiency levels. These errors are most frequent around ‘Emerging’, which requires judging whether evidence and fallacies are roughly balanced. This judgment is difficult when fallacy detection depends on fine-grained distinctions. This result can likely be attributed to the inherent difficulty of identifying specific fallacy types (e.g., causation, false equivalence). These distinctions were provided to human coders but are not explicitly available to the model [39]. Our findings suggest that while LLMs can approximate human scoring, biased judgments of the balance between evidence and fallacies underscore the need for additional scaffolding or hybrid human-AI approaches for critical thinking assessment. Future work can test whether providing explicit fallacy definitions and labeled examples improves performance.

## 5. DISCUSSION AND CONCLUSION

In this paper, we developed and evaluated automated methods for assessing critical thinking subskills in student essays, attempting to address the pressing need for more reliable and robust measures of this dynamic and understudied form of higher-order reasoning. Our work used two complementary strategies: prompting proprietary LLMs under both zero-shot and few-shot conditions and fine-tuning open-source LLMs using human-labeled data. By doing so, we investigated the capability of these systems to detect critical thinking subskills, such as constructing coherent arguments, systematically integrating evidence, and articulating meaningful counterarguments. Our aim is to provide the foundation for future EDM work on modeling critical thinking so that eventually such detectors can be used in educational environments.

Overall, these methods involve trade-offs between scoring quality and deployment constraints. Fine-tuning open-source models achieves the strongest performance, with Llama fine-tuning outperforming and ModernBERT matching, respectively, GPT-5 few-shot’s performance in Krippendorff’s α, despite their smaller size. This result highlights the value of on-task training data for critical thinking assessment. Prompting remains easier to deploy, but it can be more expensive at scale (≈ $40 for scoring 500 essays with GPT-5) and raises student data privacy concerns. At the same time, even for our best model, Llama 3.1 8B fine-tuning using rubrics and justifications, we noted inconsistent results across subskills. The model was especially challenged by subskills where there was a class imbalance and only subtle (rather than clearer) distinctions across proficiency levels. These challenges suggest that we need to incorporate data-level balancing (e.g., synthetic minority over-sampling) and cost-sensitive training into LLM fine-tuning for critical thinking assessment [59].

Our experiments with fine-tuning Llama show that adapting to unseen subskills remains difficult without on-task training data. Rubric descriptors are necessary for cross-subskill generalization, and generating justifications further improves performance. This result likely follows because justifications encourage the model to explain its label using rubric criteria and evidence from the essay, which helps it map an unseen subskill to the proficiency levels more consistently. However, reliability remains below our target threshold. Performance also drops relative to settings where the target subskill is seen during training. This finding suggests that robust scoring of new subskills will require some subskill-specific human supervision, such as a small set of labeled essays or curated in-context examples. Future work may also investigate meta-prompting methods to automatically generate rubrics given description of previously unseen skills. Deploying automated critical thinking assessment in practice will require either subskill-specific labeled data or mechanisms to encode transferable traits across related subskills.

We acknowledge that selecting 0.6 as a threshold for reliability (compared to typical 0.8) affects the interpretation and use of our findings. However, this decision was intentional given the exploratory nature of this work and our long-term goal of developing a model for critical thinking that can be used in AI tools. A limitation of this initial work is that we evaluated a small set of models and prompting settings, even as the space of LLMs and prompting strategies for AES and related tasks continues to expand. It is therefore possible that other proprietary models (o3, o4-mini-high, Opus 4.6), alternative prompting strategies, or different open-weight models (Qwen [3], or DeepSeek [12]) would yield stronger performance for critical thinking assessment. The bottom line is that our experiments provide evidence that open-weight LLMs, after some training, can be competitive in this setting. The rapid pace of LLM development remains a challenge for interpreting and comparing results over time. Similarly, a broader range of critical thinking skills and contexts for its application will need to be studied before we can confidently draw conclusions about the usefulness of LLMs for measuring critical thinking as a whole.

Nonetheless, these preliminary results demonstrate that there is potential for LLMs to assess a range of critical thinking skills efficiently, rapidly, and cost-effectively, in the context of authentic student work. LLMs can potentially become a tool to support expanded and more extensive research on critical thinking, the relationship between its components, the processes involving its development, and its relationships to other forms of complex student cognition. This work lays the groundwork for AI models that can give teachers meaningful insights into how students’ critical thinking is developing and that can be embedded in adaptive learning systems to help cultivate the skills humans will need most as they navigate an era of AI-generated content, disinformation, and increasingly complex global challenges.

## 6. DATA AVAILABILITY STATEMENT

The data and code used in this study are openly available in GitHub at https://github.com/augmented-org/critical-thinking.

# 8. ACKNOWLEDGEMENTS

The research reported here was supported by AugmentED, which is part of the of the Advanced Education Research and Development Fund (AERDF). The opinions expressed are those of the authors and do not necessarily represent views of the AERDF.